\documentclass[onecolumn,11pt,a4paper,aps]{revtex4}

\newcommand{\eq}[1]{\begin{equation}#1\end{equation}}

\usepackage{amsmath}
\usepackage{graphics}
\usepackage{graphicx}
\usepackage{psfrag}

\begin{document}

\title{On the relation between entanglement and subsystem Hamiltonians}

\author{Ingo Peschel$^1$ and Ming-Chiang Chung $^{2,3}$}
\affiliation{
$^1$Fachbereich Physik, Freie Universit\"at Berlin, Arnimallee 14, D-14195 Berlin, Germany\\
$^2$Physics Division, National Center for Theoretical Science, Hsinchu 30013, Taiwan\\
$^3$Institute of Physics, Academia Sinica, Taipei 11529, Taiwan
}

\begin{abstract}
We show that a proportionality between the entanglement Hamiltonian and the Hamiltonian
of a subsystem exists near the limit of maximal entanglement under certain conditions. 
Away from that limit, solvable models show that the coupling range differs in both 
quantities and allow to investigate the effect.
\end{abstract}

\maketitle

\vspace{1cm}

The reduced density matrix (RDM) which describes a subsystem $\alpha$ of a total system in 
a pure quantum state has been the topic of numerous studies. It can be written in the form 
$\rho_{\alpha} = exp(-\mathcal{H}_{\alpha})/Z$ with an operator $\mathcal{H}_{\alpha}$ which 
has become known as entanglement Hamiltonian \cite{Li/Haldane08}. For free fermionic or 
bosonic systems in their ground state, $\mathcal{H_{\alpha}}$ has again free-particle form 
and can be determined explicitly, see \cite{Peschel/Eisler09} for a review. Because of the 
thermal form of $\rho_{\alpha}$, the question whether $\mathcal{H_{\alpha}}$ is related to the 
subsystem Hamiltonian $H_{\alpha}$ arises naturally. From the exact results, one sees that the 
answer is in general no. For example, segments in non-critical quantum chains like
the transverse Ising model or a dimerized hopping model lead to a single-particle spectrum 
in $\mathcal{H_{\alpha}}$ which is linear near zero, whereas $H_{\alpha}$ has energy bands with 
a gap. Moreover, the low-lying eigenfunctions of $\mathcal{H_{\alpha}}$ are concentrated near 
the boundaries while they are extended in $H_{\alpha}$ except for zero-energy modes.
A certain similarity exists only in the
critical case, where both spectra are asymptotically linear with level spacing $1/\ln L$ and 
$1/L$, respectively, where $L$ is the length of the subsystem. This allows to define an
effective temperature in the RDM \cite{Eisler/Legeza/Racz06}. However, the eigenfunctions 
of the two Hamiltonians still differ, and also the forms of $\mathcal{H}_{\alpha}$ and 
$H_{\alpha}$ in real space, see \cite{Peschel/Eisler09}. 

The situation becomes different if the subsystem is translationally invariant, as is
the case for sublattices in a chain or for a leg of a ladder. Then the eigenfunctions of
$\mathcal{H_{\alpha}}$ and $H_{\alpha}$ are both momentum eigenstates and a closer relation 
is possible, although not necessary. For a transverse Ising chain, for example, the 
sublattices simply decompose into the individual sites, but $\mathcal{H}_{\alpha}$ has 
non-trivial momentum-dependent excitations in the fermionic representation 
\cite{Igloi/Peschel10}. In other cases, however, correspondences between the spectra were 
found, see e.g. \cite{Lauchli10,Schliemann11} for quantum Hall systems and \cite{Poilblanc10} 
for a Heisenberg ladder. This feature was explained in a recent paper for coupled conformally 
invariant subsystems with left- and right-moving particles \cite{Qi/Katsura/Ludwig11}. 

In the present note, we want to point out that a relation 
$\mathcal{H_{\alpha}} \sim H_{\alpha}$ can be obtained very simply via perturbation theory 
for a total system formed from two strongly coupled subsystems.
This is essentially also the case treated in \cite{Qi/Katsura/Ludwig11}. We also show,
for a solvable fermionic system, how away from strong subsystem coupling the operator
$\mathcal{H_{\alpha}}$ contains longer-range interactions, as found numerically in    
\cite{Cirac/Poilblanc/Schuch/Verstraete11} for Heisenberg and AKLT ladders.

Consider a quantum system made up of two parts with Hamiltonians $H_1$ and $H_2$ coupled
via the Hamiltonian $H'$. This could be a ladder with two legs and rungs described by $H'$.
We assume $H'$ large and treat $H_1+H_2$ as a perturbation. Then, if $|\Psi_0>$ is the 
(non-degenerate) ground state of $H'$, it changes in first order to 
\eq{
|\Psi_0^1> = |\Psi_0> - \sum_{k \neq 0} |\Psi_k> \frac{<\Psi_k|(H_1+H_2)|\Psi_0>}{E_k-E_0} 
\label{Psi1}}
where $|\Psi_k>$ are the eigenfunctions of $H'$ and $E_k$ the eigenvalues. We now assume

(1) There is only coupling to excited states with the same gap  $\Delta = E_k-E_0$

(2) Both $H_{\alpha}$ give the same matrix elements, $<\Psi_k|H_1|\Psi_0> = <\Psi_k|H_2|\Psi_0>$

Then
\eq{
|\Psi_0^1> = |\Psi_0> - \frac {2}{\Delta} \sum_{k \neq 0} |\Psi_k> <\Psi_k|H_1|\Psi_0> 
\label{Psi2}}
which can be written
\eq{
|\Psi_0^1> = |\Psi_0> - \frac {2}{\Delta} \hat{H_1}|\Psi_0>
\label{Psi3}}
where $\hat{H_1} = H_1 -<H_1>$ with $<H_1>=<\Psi_0|H_1|\Psi_0>$. The total density matrix then 
is, to first order,  
\eq{
\rho^1 = |\Psi_0><\Psi_0| - \frac {2}{\Delta} \left[\hat{H_1}\;|\Psi_0><\Psi_0| +
        |\Psi_0> <\Psi_0|\;\hat{H_1} \right]
\label{rho}}
Since $H_1$ operates only in subsystem 1, the trace over subsystem 2 can be taken and leads
to 
\eq{
\rho^1_1 = \rho_1  - \frac {2}{\Delta} (\hat{H_1} \rho_1 + \rho_1 \hat{H_1})
\label{rho11}}
where $\rho_1$ is the RDM for $|\Psi_0>$. If now $\rho_1$  is a multiple of the unit matrix,
which means that $|\Psi_0>$ is maximally entangled, it can be pulled out in front and one can 
write, exponentiating the differences
\eq{
\rho^1_1 = \frac{1}{Z} \exp(- \frac {4}{\Delta} H_1)
\label{rho12}}
where $Z=\mathrm{tr}_1(1-4 H_1/\Delta)=\mathrm{tr}_1\exp(-4 H_1/\Delta)$. Thus $\rho^1_1$ is 
correctly normalized to first order and one has the relation
\eq{
 \mathcal{H}_1 = \frac {4}{\Delta} H_1
\label{relation}}
i.e. a direct proportionality between the two Hamiltonians. The quantity $\Delta/4$ can be
viewed as an effective temperature and by assumption one is in the high-temperature limit.
Condition (2) could be weakened to a proportionality between the matrix elements. This would 
only change the prefactor in (\ref{relation}).

The conditions used in the derivation are not as restrictive as they may seem. They are fulfilled, 
for example, for an antiferromagnetic Heisenberg ladder with Hamiltonian
\eq{
H = H_1 + H_2 + H' = J \sum_n {\bf{S}}_n {\bf{S}}_{n+1}+ J \sum_n {\bf{T}}_n {\bf{T}}_{n+1}+
 J' \sum_n {\bf{S}}_n {\bf{T}}_n
\label{Heisenberg}}
where the $S$ and $T$ are spin one-half operators. Then the ground state $|\Psi_0>$ is a product of 
singlets at the different rungs. Each singlet is maximally entangled and gives a RDM which is
1/2 times the $2 \times 2$ unit matrix. Each term in $H_{\alpha}$ has matrix elements to triplet states at
two neighbouring rungs which leads to $\Delta = 2J'$. Thus the coupling in $\mathcal{H}_1$ is given 
by $K = 4J/2J' = 2J/J'$. This is exactly the result found numerically in 
\cite{Cirac/Poilblanc/Schuch/Verstraete11} in the limit $J \ll J'$, see Fig. 5(a) there. In their
notation, $K = 2 \cos(\theta)/\sin(\theta)$ and one has to consider $\theta \approx \pi/2$ where
$K = 2(\pi/2-\theta)$.

One can ask if the considerations also hold for an anisotropic Heisenberg model. If the rung
coupling remains isotropic, this is indeed the case, since the singlet-triplet level scheme for each
rung does not change. However, if $H'$ is of XXZ form, one has two single levels and one doublet.
Then there are excitations with two different gaps to the spin singlet, which is the lowest state 
throughout the planar region ($|J'_z| \le J'_x=J'_y=J')$. These appear with different pieces of $H_1$ 
and (\ref{relation}) is changed to
\eq{
 \mathcal{H}_1 = \frac {4}{\Delta_{xy}} H_{1,xy} + \frac {4}{\Delta_z} H_{1,z}
\label{relationXXZ}}
where $\Delta_{xy}=J'+J'_z$ and $\Delta_{z}=2J'$. Thus while the Heisenberg form remains, the anisotropy 
of $\mathcal{H}_1$ is not the same as that of $H_1$. An exception is the planar case, $H_{1,z}=0$. Then $H_1$
couples only to the doublet and the formula (\ref{relation}) with the proper gap holds again. This is 
interesting, because $H_1$ is then solvable in terms of fermions 
whereas the Hamiltonian of the ladder is not.

The considerations also apply to a fermionic system as treated in \cite{Qi/Katsura/Ludwig11}. 
Consider two species of fermions with opposite dispersion and mutual coupling. The Hamiltonian is
\eq{
H = H_1 + H_2 + H' = \sum_q \gamma_q\, a_q^{\dag}a_q - \sum_q \gamma_q\,b_q^{\dag}b_q + 
                      \sum_q \delta\, (a_q^{\dag}b_q+b_q^{\dag}a_q)   
\label{fermions}}
where $q$ denotes the momentum. If $\gamma_q=q$, this describes two systems with only right- or left-
moving particles. If $\gamma_q=\mathrm{cosq}$, it describes a ladder with opposite nearest-neighbour 
hopping matrix elements in the two legs. The coupling term $H'$ is diagonalized by the operators 
$(a_q \pm b_q)/\sqrt 2$ and gives the two single-particle levels $\pm \delta$ for each $q$, thus
$\Delta=2\delta$. Moreover, the levels are analogous to spin singlets and therefore
maximally entangled. The operators $H_{\alpha}$ have equal matrix elements between them. Therefore
(\ref{relation}) holds for large $\delta$ and $\mathcal{H}_1$ is of the form 
\eq{
\mathcal{H}_1 = \sum_q \varepsilon_q\, a_q^{\dag}a_q 
\label{fermions2}}
with $\varepsilon_q=4\gamma_q/\Delta = 2\gamma_q/\delta$. This is the result found in 
\cite{Qi/Katsura/Ludwig11}.

For this system, however, $\mathcal{H}_1$ can be determined exactly and the $\varepsilon$ 
follow from the eigenvalues of the correlation matrix in the subsystem 
\cite{Peschel03,Vidal03,Peschel/Eisler09,Latorre/Riera09}. But because of the translation 
invariance, this matrix is diagonal in momentum space and the eigenvalues are given by the 
occupation numbers $n_q=<a_q^{\dag}a_q>$. Diagonalizing (\ref{fermions}) with a canonical 
transformation $a_q=u_q\alpha_q+v_q\beta_q, b_q=-v_q\alpha_q+u_q\beta_q$ where $u_q^2+v_q^2=1$,
one obtains
\eq{
H =  \sum_q \omega_q (\alpha_q^{\dag}\alpha_q - \beta_q^{\dag} \beta_q) \,\, ,\hspace{1cm}
        \omega_q= \sqrt{\gamma_q^2+\delta^2}
\label{fermions3}}
This gives the occupation numbers
\eq{
n_q= v_q^2 = \frac {1}{2}(1- \frac{\gamma_q}{\omega_q}) 
\label{occupation}}
and leads to $\varepsilon_q=\ln[(1-n_q)/n_q]$, or
\eq{
\varepsilon_q =\ln \left(\frac{\omega_q+\gamma_q}{\omega_q-\gamma_q}\right)
\label{epsilon}}
In \cite{Qi/Katsura/Ludwig11} this was obtained in a different way. 
If one considers the other subsystem, $v_q^2$ is replaced by $u_q^2$, which changes the sign of
$\varepsilon_q$ but not the RDM spectrum.
 
Expanding (\ref{epsilon}) for large $\delta$, one reobtains the result $\varepsilon_q=
2\gamma_q/\delta$ found above. In the opposite case, $\delta \ll \gamma_q$, however,
the variation is logarithmic, $\varepsilon_q=2\ln(2\gamma_q/\delta)$. The variation of
$\varepsilon_q$ with $q$ for $\gamma_q=-\mathrm{cosq}$ is shown in Fig. 1 for several 
values of  $\delta$. One sees that the amplitude increases as $\delta$ becomes smaller.
At the same time, the curves deviate from a simple cosine function and become more
rectangular. This is illustrated for $\delta=0.1$ by the dotted line. Near the points 
$q=\pm \pi/2$, one is always in the strong-coupling limit and the slope is
$\pm 2/\delta$.

\pagebreak

%
\begin{figure}
\begin{center}
\includegraphics[width=9.cm,angle=0]{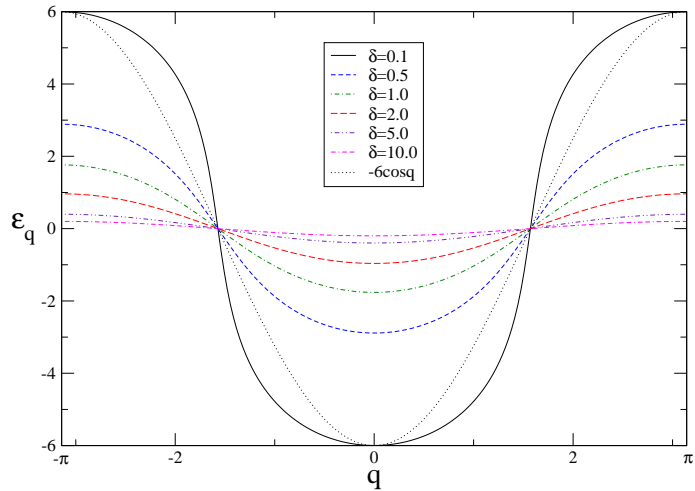}
\caption{Dispersion relation for the single-particle excitations in $\mathcal{H}_1$
for the model (\ref{fermions}) with $\gamma_q=-\mathrm{cosq}$ and several values of 
$\delta$. The dotted line shows a cosine for comparison.}
\label{fig1} 
\end{center}  
\end{figure}
%

The deviation of $\varepsilon_q$ from $\gamma_q$ means that the 
hopping in $\mathcal{H}_1$ is different from that in $H_1$. In the example, where one
has nearest-neighbour hopping in $H_1$, one finds hopping to more distant sites in 
$\mathcal{H}_1$. This can be seen directly by expanding (\ref{epsilon}) to higher orders. 
It is more instructive, though, to show the result graphically. This is done in Fig. 2, 
where the amplitudes $t_n$ for hopping to the n-th neighbour are plotted, normalized
by $t_1$. Shown are only those for odd distances, since the other ones are zero.

%
\begin{figure}
\begin{center}
\vspace{0.5cm}
\includegraphics[width=10cm,angle=0]{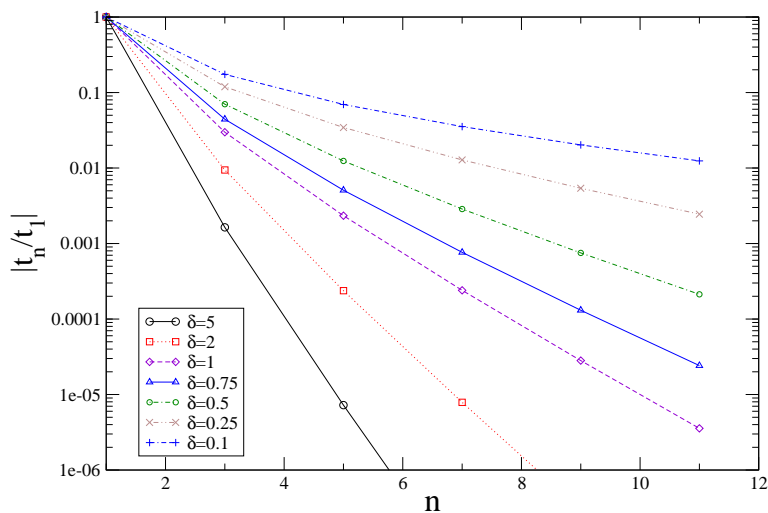}
\end{center}
\caption{Hopping amplitudes corresponding to the dispersion relations in Fig. 1.
Shown are the absolute values of $t_n/t_1$ for odd distances $n$. 
\label{fig2}
}   
\end{figure}
%
For large $\delta$, one sees a very rapid decrease of the $t_n$ with $n$ and only $t_1$
is relevant. As $\delta$ decreases, the decay slows down and also longer-range
hopping becomes becomes important. Due to the shape of $\varepsilon_q$, however, the 
dominant term is always $t_1$. Very similar results were obtained numerically in
\cite{Cirac/Poilblanc/Schuch/Verstraete11} for Heisenberg ladders. Formally, they are connected
with the higher orders in the perturbation expansion for $|\Psi_0>$. The effect is reminiscent
of the situation for transfer matrices in two-dimensional Ising or Gaussian models, where 
the exact operators in the exponent and those of the Hamiltonian limit differ in dispersion
relation and coupling range.

Summing up, we have shown how a proportionality between $\mathcal{H}_1$ and $H_1$ can
be obtained for strongly coupled and maximally entangled subsystems by treating the
subsystem Hamiltonians in first-order perturbation theory. Whether they describe a critical 
or a non-critical system does not matter, only their smallness enters. The entanglement
is decresed only weakly in this case. The fermionic example showed explicitly how
the situation changes away from the strong-coupling limit. In the free-fermion and
free-boson case, one can find a number of simple systems, where an expression as   
(\ref{epsilon}) appears. Examples are hopping ladders with alternating rung couplings or
the BCS model and the (bosonic) Luttinger model as systems of right- and left-moving 
particles. An exception is a homogeneous hopping ladder. There $H_1+H_2$ commutes with 
$H'$ and does not change the wave function. The matrix elements in condition (2) are
then of opposite sign.

Finally, one should mention that formulae similar to (\ref{occupation}), (\ref{epsilon})
have appeared before in studies of quenches in quantum chains 
\cite{Rigol/Muramatsu/Olshanii06,CC07}. In this case, one determines the occupation numbers 
for the modes of the new Hamiltonian in the state before the quench, using the appropriate 
canonical transformation. If, for example, one starts in the ground state of a hopping chain 
with alternating site energies $\pm \delta$ and switches this dimerization off, as done in 
\cite{Rigol/Muramatsu/Olshanii06}, the $n_q$ of the modes in the homogeneous chain are exactly 
(\ref{occupation}) with $\gamma_q=\mathrm{cosq}$. As above, one can then define a thermal 
density matrix and an effective Hamiltonian, and the difference is only that these quantities 
refer to the ${\it{full}}$ system and not to a part of it. The relation  
$\varepsilon_q = 2\gamma_q/\delta$ in this case was already noted in \cite{Peschel/Eisler09}.

\end{document}